\documentclass[aps,preprint,nofootinbib,superscriptaddress]{revtex4}
\usepackage{graphicx,color}
\usepackage{amsmath,amssymb,amscd,amsfonts}
\usepackage{subfigure}
\usepackage{slashed}
\usepackage{bm}
\usepackage{xcolor}
\usepackage{diagbox,pict2e}

\def \Lag{\mathcal L}

\newcommand{\bwt}{\begin{widetext}}
\newcommand{\ewt}{\end{widetext}}
\newcommand{\newc}{\newcommand}
\newc{\hc}{\dagger}
\newc{\pd}{\partial}
\newc{\beq}{\begin{equation}}
\newc{\eeq}{\end{equation}}
\newc{\beqa}{\begin{eqnarray}}
\newc{\eeqa}{\end{eqnarray}}
\newc{\bi}{\begin{itemize}}
\newc{\ei}{\end{itemize}}
\newc{\ra}{\rightarrow}
\newc{\la}{\leftarrow}
\newc{\lra}{\longrightarrow}
\newc{\lla}{\longleftarrow}
\newc{\Lra}{\Longrightarrow}
\newc{\Lla}{\Longleftarrow}
\newc{\half}{\frac{1}{2}}
\newc{\fth}{\frac{1}{4}}
\newc{\del}{\delta}
\newc{\Del}{\Delta}
\newc{\gm}{\gamma}
\newc{\Gm}{\Gamma}
\newc{\lam}{\lambda}
\newc{\kap}{\kappa}
\newc{\tri}{\triangle}
\newc{\eps}{\epsilon}
\newc{\epsp}{\epsilon^\prime}
\newc{\veps}{\varepsilon}
\newc{\wt}{\widetilde}
\newc{\ovl}{\overline}
\newc{\tchi}{\tilde{\chi}}
\newc{\ds}{\displaystyle}
\newc{\pmt}{\pm\!\pm}
\newc{\PL}{\hat{L}}
\newc{\PR}{\hat{R}}
\newc{\st}{s_\theta}
\newc{\ct}{c_\theta}

\newc{\msm}{\mathrm{SM}}
\newc{\msh}{\mathrm{sh}}
\newc{\mtev}{\mathrm{TeV}}
\newc{\mgev}{\mathrm{GeV}}
\newc{\mmev}{\mathrm{MeV}}
\newc{\mkev}{\mathrm{keV}}
\newc{\mev}{\mathrm{eV}}
\newc{\Tr}{\mathrm{Tr}}
\newc{\non}{\nonumber}
\newc{\NLDBD}{$0\nu\beta\beta \,$}
\newc{\NLDBDs}{$0\nu\beta\beta$s}
\newc{\bfml}{\mathbf{m_L}}
\newc{\bfmr}{\mathbf{m_R}}
\newc{\bfmd}{\mathbf{m_D}}
\newc{\clbl}{\color{blue}}
\newc{\clg}{\color{green}}
\newc{\clr}{\color{red}}
\mathchardef\mhyphen="2D
\newc{\SL}{\not\!\!}
\newc{\loe}{\frac{L}{2E}}

\begin{document}
\title{Consequences of Neutrinoless Double Decays Dominated by Short Ranged Interactions }
\date{\today}

\author{C.Q. Geng}
\email{geng@phys.nthu.edu.tw}
\affiliation{School of Fundamental Physics and Mathematical Sciences\\Hangzhou Institute for Advanced Study, UCAS, Hangzhou 310024, China}
\affiliation{International Centre for Theoretical Physics Asia-Pacific, Beijing/Hangzhou, China}
\affiliation{Department of Physics, National Tsing Hua University,
Hsinchu, Taiwan 300}
 \affiliation{National Center for Theoretical Sciences, Hsinchu,
Taiwan 300}

\author{John N. Ng}
\email{misery@triumf.ca}
\affiliation{TRIUMF, 4004 Wesbrook Mall, Vancouver, BC V6T 2A3, Canada}
\begin{abstract}
We investigate some consequences if neutrinoless double beta decays (\NLDBDs) of nuclei are dominated by short range interactions.
To illustrate our results, we assume that \NLDBDs proceed mainly through short range interactions involving  two-W-boson exchanges
and confine ourselves to only include new scalars without new gauge interactions for the SM fermions.
For the neutrino mass problem, we propose to solve it by adopting that the light neutrinos have predominantly
Dirac masses and the small Majorana masses induced by the new scalars render them quasi-Dirac
particles. 
This particular aspect of neutrinos may be detectable in the next generations of
neutrino oscillation experiments and/or neutrino telescope. If so this opens a new connection between
\NLDBD and neutrino physics.
We also noted the new physics signals such as the high charged states that can be explored in hadron colliders.
In particular, we find that a high energy $e^-e^-$ will be very useful in testing the origin of lepton number violation, which complements
the \NLDBD studies.

\end{abstract}
\maketitle

\section{Introduction}

In this paper we investigate the possibility that neutrinoless double beta decays (\NLDBDs) of nuclei are
dominated by short range physics not involving a heavy sterile right-handed neutrino but due to some other new physics beyond the Standard Model (SM).
This is in sharp contrast to the usual assumption that \NLDBD  is due to the
exchanges of light Majorana neutrinos, which constitute a long range exchange force between decaying nucleons in the
nucleus. The half-life of the decaying nucleus is directly proportional to the masses of the exchange
or virtual neutrinos. They are identified as the active neutrinos of the SM, which are known to be massive but
light, i.e. less than 1 eV, due to
the observed neutrino oscillations. We refer this as the three-Majorana-neutrino paradigm (3MNP). This is an economical
and elegant framework for \NLDBD as it involves only  physics in the SM with the violation of lepton number encoded in
the Majorana masses of the active neutrinos. For an up to date review, see \cite{DPR,BG}. However, since the origin
of the active neutrinos masses and their nature is an open question, it behoves us to examine alternatives to
the above paradigm and to study consequences that they will lead to, in particular to search
for new pathways that are not evident in the 3MNP.

The scenario we are interested in assumes that \NLDBDs  proceed predominately by short range physics beyond the SM. This can happen if all the active neutrinos have physical masses too small to induce \NLDBD even if they were Majorana particles. A less stringent possibility is that the Majorana phases are such that they cancel in
the effective $\nu_e$ Majorana mass. This amounts to the first element $M_\nu$ of the active neutrino mass matrix in the weak basis is vanishing,
i.e. $(m_L)_{ ee}\simeq 0$. A third possibility is that the three active neutrinos participating
in oscillations have dominantly Dirac masses. If any of the above scenarios takes place and \NLDBD is observed
in the next generations of experiments then it is likely that some new short range physics is operative. Short range physics contributions to \NLDBDs  have been discussed in \cite{GDIK,CDVGM} concentrating in how they affect the hadronic physics. They have been parametrized
by effective operators of dimension 7 and 9 \cite{PHKK99,PHKK01}. Since dimension 7 operators will involve a light neutrino exchange, they will fall out of our assumptions. That leaves dimension 9 as the lowest dimensional
operator we need to consider. The theory space for new physics that can generate these operators is large.
To reduce that we make a conservative assumption that all SM fermions do not carry additional quantum numbers than dictated by the SM gauge symmetry. This is supported by LHC having not seen any new gauge bosons and numerous low energy precision measurements that set stringent limits on their masses and couplings. This leaves new scalars and fermions transform nontrivially under the SM gauge symmetry that can carry color and color singlets as new degrees of freedom to be studied. Here, we shall concentrate on scalars and leave new fermions for a future work.

We organize our paper as follows. In Sec.~2, we take two-W-boson exchange as the lowest state for the dimension 9
operator. Then  new physics for \NLDBD will proceed via $WW\ra ee$. Tree level new physics for this
will involve colorless scalars. We then compare the constraints given by \NLDBD with that from the LHC and future colliders on these new objects. Since the interactions involve must violate lepton number by 2 units,
one has to check that they do not generate $(m_L)_{ee}$ at a large enough value so as to invalid our short range
dominance proposition. This is independent of the black box theorem \cite{SV}, which generates a Majorana
mass for $\nu_e$ at the 4-loop level. Quantitatively, this yields $(m_L)_{ ee}\lesssim 10^{-28}
$ eV \cite{MLM}. This value is inconsequential for \NLDBD  if they were to be discovered in the current or next generation experiments.
In Sec.~3,  we investigate the issue of neutrino masses if \NLDBD is driven by
the short range interactions proposed. Our conclusions
are given in Sec.~4.
\section{2-W-bosons mechanism for \NLDBD}

\begin{figure}[h!]
\centering
\includegraphics[width=0.8\textwidth]{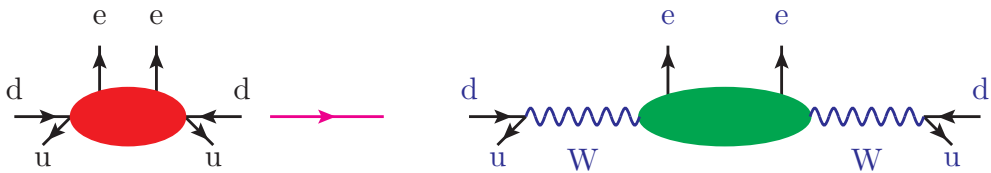}
\caption{Feynman diagrams for the short range interaction of neutrinoless double beta decays,
where the righthand side indicates the 2-W-boson exchange mechanism.}
\label{fig:0nubb}
\end{figure}

At the quark level, \NLDBD can be represented by Fig.~\ref{fig:0nubb} with the 2-W-boson mechanism
being the leading approximation as depicted.

The Lagrangian for the short range interaction can be symbolically written as
\beq
\Lag=\frac{G_F^2}{2m_p} \sum_{i} \epsilon_i J_i J_i j_i,
\label{eq:lag}
\eeq
where $i$ denotes different electron currents ($j_i$) and quark currents ($J_i$), $\epsilon_i$ represents the particle physics involved,  and $m_p$ stands for
the proton mass. Here, Lorentz contractions have to be taken and are not shown, while the quark currents are to be sandwiched between initial and final
nuclear states for the full matrix element.  The half-life for a given decay may be generically expressed as
\beq
\label{eq:half}
T_\half^{-1}= |\eps_i|^2 G_i \left|{\mathcal{M}_i}\right|^2,
\eeq
where $G_i$ is the nuclear phase space factor, and ${\mathcal{M}_i}$. The calculations of these two factors
are given in \cite{GDIK}. In this paper, we are concerned with models that give rise to $\eps_i$ and it is
dimensionless. As a comparison, the long range neutrino exchange is given by $\eps_\nu = (m_L)_{ ee}/(0.01 \mathrm{eV})$.

\begin{figure}[h!]
\centering
\includegraphics[width=0.4\textwidth]{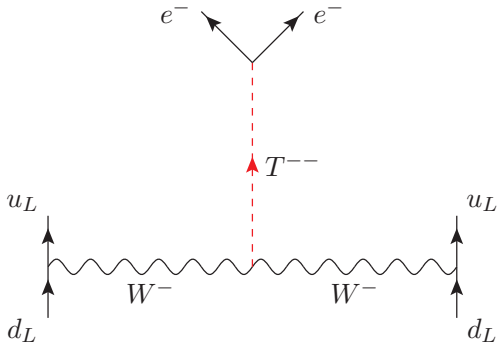}
\caption{Generic doubly charged scalar exchange for neutrinoless double beta decays}
\label{wwgen:fig}
\end{figure}

For the 2-W-boson scattering mechanism one can construct tree level new physics that induces $W^{-*}+W^{-*}\ra e^- + e^-$.
Immediately one can recognize that a doubly charged scalar $T^{--}$ will be involved.  A generic $T^{--}$ exchange mechanism
is displayed in Fig.~\ref {wwgen:fig}.
The nature of $T^{--}$ depends on whether the chirality of the final state electrons.
A detail examination of this is given next.

\subsection{$W^{-*}+W^{-*}\ra e_L +e_L$}

If the electrons are left-handed they are in a SM doublet with the active neutrinos written as $\ell=(e_L \nu_L)$. Then $T$ will be in a $SU(2)_L$ triplet with
hypercharge $Y=1$ where we use the normalization $Q=T_3+Y$ with standard notations. Explicitly, $T$ consists of three states: $(T^{++},T^+,T^0)$. A coupling between $T$ and the lepton, given by $y_\ell \bar{\ell^c}T \ell$, can be constructed. This type of Higgs triplet models is popular in type II and radiative seesaw models for neutrino masses, see e.g. \cite{CGN,CGNW}.
 The $T^0$ component must pick up a VEV, $v_T$, in order to get a $W^{-} W^{-}T^{++}$ coupling.
 This in turn generates a tree level neutrino Majorana mass; hence $y_\ell v_T < 1\mathrm{eV}$. Thus, the $WWT$ coupling is $\sim gv_T$, where $g$ is the $SU(2)$ gauge coupling.
 Since no doubly charged scalar has been found at the LHC~\cite{Atlasch,CMSch},  a lower bound on the mass is 1.3 TeV
 by using the signal of same sign dileptons
and assuming 100\% branching ratio. The rate for \NLDBD is given by $g y_\ell v_T m_p/M_T^2$.
 We estimate that $\eps \lesssim 10^{-24}$ and so this case is uninteresting. We conclude that triplet Higgs
 with SM quantum numbers $(1,3,1)$ in usual notations will not be useful for us.

 \subsection{$W^{-*}+W^{-*}\ra e_R +e_R$}

 The electron pairs are in a singlet state of $(1,\mathbf{1},-2)$ under $SU(3)_C\times SU(2)_L\times U(1)_Y$.
 A doubly charged scalar singlet $\Phi (1,\mathbf{1},2)$
 can couple to them with the coupling $1/2y_\Phi \bar{e_R^c}e_R \Phi$. Now $y_\Phi$ is unconstrained by neutrino
 masses. On the other hand, $\Phi$ will have no tree level coupling to the two W-bosons. This necessitates the
 introduction of additional Higgs scalars. We have previously ruled out the scalar $(1,\mathbf{3},1)$ and it is easy to see that triplets with $|Y|\geq 2$ cannot be used. This leaves the option of higher $SU(2)$ representations.
 The next lowest representation that can be used is $\Psi(1,\mathbf{4},3/2)$ and explicitly the quadruplet states are
 $(\psi^{+++},\psi^{++}/\sqrt{3},\psi^{+}/\sqrt{3},\psi^{0})$. If $\langle \psi^0 \rangle =v_\psi\neq 0$ the vertex $W^-W^- \psi^{++}$ can be generated and is $igv_\psi/2.$. The hypercharge assignments also prevents tree level couplings to active neutrinos.

\begin{figure}[h!]
 \centering
 \includegraphics[width=0.4 \textwidth]{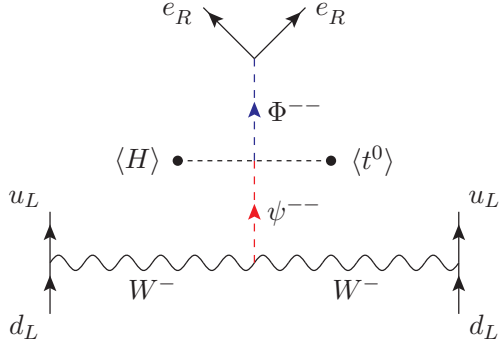}
 \caption{Neutrinoless double beta decays via the doubly charged Higgs exchange in the weak interaction basis.}
 \label{psiphi:fig}
 \end{figure}

 The next ingredient is to provide the mixing between $\Psi^{++}$ and $\Phi^{++}$. The price to pay is to introduce yet another scalar $T^\prime (1,\mathbf{3},0)=(t^{-},t^{0}/\sqrt{2},t^{+})$. Then the gauge invariant term
 $ H T^\prime \Psi \Phi^{\dag}$ is allowed \footnote{If economy on new states is desired one can construct a soft
 term such as $\Psi\Psi \Phi^{\dag}$. Mixing between $\Psi$ and $\Phi$ is induced after the spontaneous symmetry breaking (SSB) of $\Psi$. However, this will require that $\Psi$ be in an odd dimensional $SU(2)$ representation. For our case the lowest one would be a quintuplet \cite{CGH}.}. We obtain the desired mixing if $\langle t^0 \rangle$ is non-vanishing
 and $\langle H \rangle = v$. We note that both $\langle t^{0} \rangle$ and $\langle \psi^0 \rangle$ must be
 less than a GeV from precision electroweak measurements. Moreover, without exotic fermions
 the hypercharges  of the new scalars are such that they have no tree level couplings to active neutrinos.
 Thus, \NLDBD is given by Fig.~\ref{psiphi:fig}.
 Our solution is not unique and higher $SU(2)$ representations can be used. Constructing viable models can simply follow what we have presented.

We return to the discussion of our model. Firstly, we can identify the origin of lepton number violation.
It is the four scalar term $HT^\prime\Psi\Phi^{\dag}$ after the SSB of the Higgs
fields. A mixing of $\Phi$ and  $\Psi$ is then generated. Explicitly, one has that
\beq
\label{eq:SSB}
\lambda H T^\prime \Psi \Phi \xrightarrow{\mathrm{SSB}}\frac{v_T v}{\sqrt{6}}\psi^{++}\Phi^{++},
\eeq
where  $v$ is the SM Higgs VEV and $\lambda$ is a free
parameter that controls the strength of lepton number breaking.  The physical doubly charged scalars
can be obtained by diagonalizing a $2\times 2$ mass matrix with off diagonal terms given by Eq.~(\ref{eq:SSB}).
The details depend on the scalar potential and are not essential for this discussion. It suffices to know that
 the mixing angle
 $\alpha$ is given by
 \beq
 \label{eq:scalarmix}
 \sin 2\alpha \sim \frac{\lambda v_T v}{M_\Psi^2- M_\Phi^2},
 \eeq
where $M_\Psi$ and $M_\Phi$ are the masses of the respective fields before diagonalisation and we have omitted some unimportant constants.  $v_T$ is constrained to be less than a few GeV,  and the masses in the denominator
are of order 0.5 TeV.  This is the lower bound from \cite{Atlasch} when the branching ratio of a doubly charged scalar into a given same sign dilepton pair is 10 \%. This is more appropriate for us since we expect
$\Phi \ell \ell^\prime$ to be approximately equal. This gives 9 such decays. If the mass splitting is also
of order 100 GeV, we expect $\alpha \lesssim 10^{-3}$ if $\lambda\sim 1$.  A larger mixing can occur if the masses are accidentally degenerate or $\lambda \sim 10$. The physical states denoted by $S^{\pm\pm}_{1,2}$
are related to the weak states $\Phi^{\pm\pm}$ and $ \psi^{\pm\pm}$ via
\beqa
\label{eq:S12}
\Phi &=& \cos\alpha S_1 +\sin\alpha S_2,
\nonumber \\
\Psi &=& -\sin\alpha S_1 + \cos \alpha S_2.
\eeqa
Without lost of generality, we assume that $S_1$ is the lighter state. As we shall see later,
the mixing is small and $S_1$ is mostly $\Phi$. The masses are denotes by $M_{1,2}$, respectively.

Referring to Eq.(\ref{eq:lag}), our model gives only one
contribution and the current correlation has the form $J^\mu J_\mu j$ where $J^\mu=\bar{u}\gamma^\mu (1-\gamma^5)d$ and  $j=\bar{e^c}(1+\gamma^5 )e$. Using $\bar{y}_\Phi = y_\Phi/g$, $\eps$ is given by
\beq
\label{eq:epsm}
|\eps|\sim m_p v_\psi \sin 2\alpha \bar{y}_\Phi \biggl( \frac{1}{M_{1}^2}-\frac{1}{M_{2}^2}\biggr).
\eeq
It is sensitive to the difference of the inverse mass squared $\Del^2= 1/M_{1}^2-1/M_{2}^2$.
Currently, the half-life of the decay $^{136} \mathrm{Xe} \ra\, ^{136}\mathrm{Ba}\, e^- e^-$ \cite{KamL} gives the most stringent limit \cite{GDIK}\footnote{ Other experiments include $^{130}\mathrm{Te}$\cite{cuore} and
$^{76}\mathrm{Ge}$ \cite{gerda} decays. They give a factor of 2 to 5 less stringent limit on $\eps$.}
\beq
\label{eq:epsl}
\eps \lesssim 5\times 10^{-9}.
\eeq
Thus, we get
\beq
\label{eq:yphil}
\frac{y_\Phi}{g} \leq 1.25 \biggl(\frac{.001}{\alpha}\biggr)\biggl(\frac{1\mathrm{GeV}}{v_\psi} \biggr)\biggl(\frac{\sqrt{\Del^2}}{\mathrm{0.5 TeV}}\biggr)^2.
\eeq
This shows the complementarity of \NLDBD to direct searches at the LHC.  The direct search is sensitive to
one states at a time and depends on the decay products of the state in question due to experimental constraints.
 If $y_\Phi/g\simeq 1$, it implies that
the branching ratio of decay of the doubly charged scalar into same sign dilepton pair is not negligible
compare to that into a pair of same sign W-bosons. This is in fact the preferred search mode at the LHC.
This holds true if the mixing is small as we argue. However, if the scalars are more degenerate and the mixing
becomes large, then the gauge bosons decays can become more important. This is more challenging experimentally
but important to test the physics involved and must not be ignored.

\section{ Neutrino mass generation}

\begin{figure}
\centering
\includegraphics[width=0.4\textwidth]{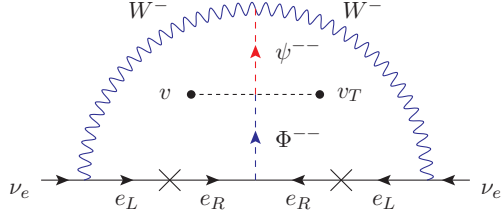}
\caption{2-loop diagram for $\nu_e$ Majorana mass }
\label{2loopnu:fig}
\end{figure}

An examination of Fig.~\ref{psiphi:fig} will show that the lepton number violating interaction
constructed will yield a 2-loop contribution to a Majorana mass to $\nu_e$. The Feynmann
diagram is given in Fig.~\ref{2loopnu:fig}, where it is depicted in the weak basis.
Evaluating the diagram \cite{CGN} for a given physical scalar $S$ gives
\beq
\label{eq:mee}
(m_L)_{ee}=g^4 m_e^2 v_\psi \overline{y_\Phi}\sin {2\alpha}\,\bigl[I(M_W^2,M_1^2,m_e^2)-I(M_W^2,M_2^2,m_e^2)\bigr].
\eeq
 The integral
I is given by
\beq
\label{eq:integral}
\begin{split}
I(M_W^2,M_S^2,m_e^2)&=\int \frac{d^4q}{(2\pi)^4} \int \frac{d^4k}{(2\pi)^4}\frac{1}{k^2-m_e^2}\\
&\times \frac{1}{k^2-M_W^2} \frac{1}{q^2-M_W^2}\\
&\times \frac{1}{q^2-m_e^2}\frac{1}{(k-q)^2-M_s^2}\,.\\
\end{split}
\eeq
A similar cancelation between $S_1$ and $S_2$ takes place as in \NLDBD. Assuming  $M_2\gg M_1 \gg M_W$, one has that
\beq
\label{eq:isim}
I(M_W^2,M_1^2,0,0)\sim \frac{1}{(4\pi)^4} \frac{1}{M_1^2}\ln^2 \bigg( \frac{M_W^2}{M_1^2}\bigg).
\eeq
Comparing Eqs.~(\ref{eq:epsm}), (\ref{eq:mee}) and (\ref{eq:isim}),  we see that $(m_L)_{ee}$ is completely determined by the amplitude for \NLDBD. There are three suppression factors at play. Firstly, the  mixing $\alpha$ is small, $10^{-3}$, secondly there is  the 2-loop
factor which is $\sim 10^{-5}$ and finally the smallness of $m_e$. The last one is the biggest suppression because $m_e^2/M_S^2 \sim 10^{-12}$ for a TeV scalar.
Even with $v_\psi \sim 1$ GeV, we get $M_{ee}\lesssim 10^{-12}\, {\mathrm{to}} \,10^{-11} \mev.$ Hence, the active $\nu_e$
will give a negligible contribution to \NLDBD and is consistent with our short range physics dominance
hypothesis. Nevertheless, we need to look deeper into what features of neutrino physics the model
will predict. To this end it is sufficient to examine the flavor diagonal elements of the active
neutrino mass it generates; i.e. the $M_{ee},M_{\mu\mu},M_{\tau\tau}$ entries. From Fig.~\ref{2loopnu:fig} and
Eq.~(\ref{eq:mee}), it is easy to see that the largest element of the $3\times 3$ neutrino mass matrix is
the $\tau\tau$ entry. We estimate that
\beqa
\label{eq:mtautau}
(m_L)_{\mu\mu}&=&\frac{m_\mu^2}{m_e^2} (m_L)_{ee}=4.3\times 10^{4}(m_L)_{ee}\sim 10^{-7}\,\mev, \\
(m_L)_{\tau\tau}&=& \frac{m_\tau^2}{m_e^2} (m_L)_{ee}=1.2\times 10^{7}(m_L)_{ee}\sim 10^{-4}\,\mev,
\eeqa
where we have assumed  that the Yukawa couplings of $\Phi$ to $e,\mu,\tau$ are the same. Since the neutrinos
oscillation data involve the mass squared differences of the neutrinos, it is obvious that
we are many orders of magnitude away from explaining the data \cite{nu2018}. The smallest
mass squared difference is $7.39 \times 10^{-5}\, \mev^2$.
Pushing some  Yukawa couplings to
their perturbative limits will not change the above conclusion.

A solution to this conundrum will
be to introduce 3 SM singlet right-handed neutrinos $N_R$ that can be used to give Dirac
masses to the active neutrinos. The physics here can be seen simply by considering the single family case with only $\nu_{eL}$ and only one $N_R$. Since we assume that lepton number violation occurs only in the scalar sector, we  take $N_R$ to have  a vanishing
tree level Majorana mass. $\nu_e$ develops a  relatively large Dirac mass via the usual Yukawa coupling
of $y_\nu \overline{\nu}_L N_R H$. If $y_\nu \sim 10^{-12}$, then a Dirac mass $m_D \sim 0.1\, \mev$
is generated for $\nu_e$ after the SSB. We treat all Yukawa couplings as free parameters and will
not go into a deeper understanding of the hierarchy shown in known Yukawa couplings. Moreover,
such a small Yukawa coupling can be implemented in extra dimensional models \cite{CNW2}

For typographic simplicity, we drop all subscripts for the lepton states in the following. The $2\times 2$ mass matrix in the $\nu,N^c$ basis is
represented by
\beq
\label{eq:nuN}
M_{\nu} = \begin{pmatrix}m_{ee}&m_D\\m_D&0 \end{pmatrix}.
\eeq
and $m_D\gg m_{ee}$ as seen previously.
We have set the lower left corner value to zero but it can be generated at 3-loop in our model.
Since this has a very small value comparing to even $m_{ee}$ we can safely set it to zero. The eigenvalues for Eq.~(\ref{eq:nuN})
are
\beqa
\label{eq:m12}
m_{\pm}&\simeq& m_D(1\pm \delta),\non \\
\delta&=&\frac{m_{ee}}{2 m_D},
\eeqa
and the states are almost maximally mixed, i.e.
\beqa
\label{eq: nustate}
\nu_+ &\simeq& \frac{1}{\sqrt 2} [(1+\theta)\nu + (1-\theta)N^c] ,\non \\
\nu_{-}&\simeq& \frac{i}{\sqrt 2} [(-1+\theta)\nu +(1+\theta)N^c],
\eeqa
with the small mixing given by $\theta=m_{ee}/(4 m_D)$.
The mass eigenstates
are a pair of Majorana leptons with opposite CP phases with a very small mass splitting, which in our example
is proportional to $m_{ee} \sim 10^{-11}\mev$.
This is known as quasi or pseudo-Dirac neutrinos \cite{wolf,petcov}.\footnote{In the usual discussion the $2\times 2$ mass matrix (see Eq.(\ref{eq:nuN}))
has 0 for the upper
 left corner and the lower right corner given by $m_R\neq 0$\cite{dGHJ}. $m_R\gg v$ is the celebrated Type I seesaw mechanism. Since most
 the signatures for psuedo-Dirac neutrino involve detecting mass splittings one
 cannot distinguish this from our scenario.}

Generalization to 3 families of neutrinos is straightforward but nontrivial. Firstly, three righthanded  SM
singlet neutrinos $N_{\alpha R}$ are introduced and denote their masses by the matrix $(\bfmr)_\alpha$, where $\alpha$
denotes weak eigenbasis. Explicitly, the weak eigenstates are
\beq
\label{eq:wkstates}
\psi_L = \begin{pmatrix} \nu_{\alpha L}\\ N_{\alpha R}^c \end{pmatrix}, \quad \quad (\alpha=e,\mu,\tau)
\eeq
where the superscript $c$ denotes charge conjugation. The neutrino mass matrix is now a $6\times 6$ matrix denoted by
\beq
\label{numm}
M=\begin{pmatrix}\bfml & \bfmd \\ \bfmd^\dagger & \bfmr\end{pmatrix},
\eeq
where  each entry $\mathbf m$ is a $3\times 3$ matrix. In our model $(\bfml)_{\alpha\beta}$ can be obtained
by calculating similar diagrams of Fig.~{\ref{2loopnu:fig}. $\bfmd$ is an obvious generalization
of the Dirac mass and is the dominant matrix, i.e. all the elements are such that
$m_D \gg m_L\gg m_R\simeq 0$. It is more convenient to diagonalize the product $M^\dagger M$,
which reads as
\beq
M^\dagger M \simeq \begin{pmatrix} \bfmd^\dagger \bfmd & \bfml^\dagger \bfmd \\ \bfmd^\dagger \bfml & \bfmd^\dagger \bfmd\end{pmatrix}.
\eeq
where the dominance of $\bfmd$ has been employed. This can be diagonalized by \cite{KL}
\beq
\label{eq: VKL}
V=\frac{1}{\sqrt 2}\begin{pmatrix} U& i U \\ U_R P& -iU_R P\end{pmatrix}.
\eeq
$U$ is the usual Pontecorvo-Maki-Nakagawa-Sakata matrix \cite{Pont,MNS} and it renders $\bfmd^\dagger \bfmd$
diagonal with the eigenvalues $m^2_1,m^2_2,m^2_3$. Defining  $\veps_i= (U^\dagger \bfml U)_{ii})$
 and $P$ a diagonal phase matrix $e^{i \phi_j}=\veps_j/|\veps_j|$. The mass eigenvalues
 are
 \beqa
 \label{eq:mvalues}
 m_{i}^{+2}&=& m_i^2 +m_i |\veps_i| ,\non \\
 m_{i}^{-2}&=& m_i^2-m_i|\veps_i|,
 \eeqa
with $i=1,2,3$. Clearly, $U$ will not diagonalize $\bfml$ in general and also $\bfmd$ is diagonalized by
$U^\dagger \bfmd U_R$ as for SM charged leptons. We label the mass eigenstates $\nu^{\pm}_{j}$
corresponding to the eigenvalues of Eq.(\ref{eq:mvalues}) and as a result the three active neutrinos
are related to the mass eigenstates via
\beq
\label{eq:wktom}
\nu_{\alpha L} =\frac{1}{\sqrt 2}\sum_j U_{\alpha j}(\nu^+_{j}+i\nu^-_{j}).
\eeq
As we have argued before the largest element in $\bfml$ is the $\tau\tau$ component; thus, we predict the ratios
$\veps_1:\veps_2: \veps_3\simeq |U_{1\tau}|^2:|U_{2\tau}|^2:|U_{3\tau}|^2\,\simeq 0.04:1:1$
with the current neutrino oscillation data given by \cite{nu2018}.

The neutrinos flavor conversion probability can be expressed as
\beq
\label{eq:nufcp}
P(\nu_\alpha\ra \nu_\beta)=\frac{1}{4}\bigg|\sum_{j=1}^3 U_{\alpha j}\{e^{-i(m_j^{+2})\loe}+e^{-i(m^{-2}_j)\loe}\}U^*_{\beta j}\bigg|^2,
\eeq
where $L$ is the baseline of the neutrino experiment and $E$ is the neutrino energy.
The $\nu_\alpha$ survival probability is then
\beq
\label{eq:paa}
\begin{split}
P(\nu_\alpha \ra \nu_\alpha)=&\sum_{j=1}^3\big|U_{\alpha j}\big|^4 \cos^2{m_j \veps_j x}\\
&+2\sum_{i>j,1}^3\big|U_{\alpha  i}\big\|^2\big|U_{\alpha j}\big|^2\cos(m_i\veps_i x)\cos(m_j\veps_j x) \cos{[(m_i^{+2}-m_j^{+2})x]},
\end{split}
\eeq
where$x=L/(2E)$ and $m_i^{+2}$ is given in Eq.(\ref{eq:mvalues}). In the limit all $\veps \ra 0$
this reduces to the standard expressions.

Eq.(\ref{eq:paa}) shows that there are long wavelength oscillations superimposed on the observed
ones.
In order to be able to observe the effects of $\veps$,  the oscillation length is given by
\beq
\label{eq:osl}
\ell= 125 E(\mmev)\bigg(\frac{10^{-5}\mev}{m\veps}\bigg) \mathrm {km}.
\eeq
The next generation reactor experiment JUNO \cite{JUNO} with a base line of 57 km and
neutrino energy in the MeV range is well suited for studying pseudo-Dirac neutrino oscillations
with splittings $O(10^{-4}) \mev$. Smaller splittings will require astrophysical neutrinos sources
and neutrino telescopes. We defer a detail study of this intricate oscillation phenomena to a future
study. For some early discussions of
the pseudo-Dirac neutrino phenomenology, see \cite{dGHJ,SP,esma}.

In conclusion by assuming the short range interactions to be dominated,
 we have broken the connection between \NLDBD and direct neutrino mass measurements using kinematics of weak decays of nuclei such as the
Katrin experiment~\cite{Katrin} and  Project 8 \cite{proj8}. This is not surprising since the neutrino exchange
is no longer assumed. If future experiments do not confirm the expected connections within expected uncertainties,
 then  short range interactions must be taken into account.
Interestingly, our study  has opened up a new connection between \NLDBD and the
phenomenology of quasi-Dirac neutrinos. These effects may be probed in future neutrino oscillation
experiments and neutrino telescopes and further studies are warranted.

\section{High Energy Collider Probes}

 To implement the short range dominance in \NLDBD,  we have introduced a moderate number of new scalars. There are a pair of neutral spin-0 states,
 $\psi^0$ and $t^0$. The real parts of which
are two Higgs scalars with masses in the Tev range. There is also one heavy pseudoscalar from a linear
combination of the imaginary parts. The orthogonal combination will be a massless Majoron since
lepton number is spontaneously violated in the model. This can serve as a candidate for dark radiation
and the phenomenology has been extensively discussed in the literature~\cite{CNW,CN1,CN2}.
In addition, there are two pairs of singly
charged scalars $(t^\pm, \psi^\pm)$ and two pairs of doubly charged scalars $(\Phi^{\pm\pm},\psi^{\pm\pm})$
and a triply charged pair $\psi^{\pm\pm\pm}$. Their masses  are all expected to be in the TeV range.
Of all these the experimentally more spectacular ones are the multiply charged states. They
are easily produced with sufficient energy. Their production cross sections are enhanced due
 to the high charges. The LHC search efforts concentrate mainly
on the doubly charged ones using multileptons as signatures which we have already discussed.
The triply charged states are more unusual can also be searched for at the LHC \cite{CGH}.
Much of the detail phenomenology is model dependent and in particular is sensitive to the parameters
of the scalar potential. Instead we will focus on more
model independent signatures without having to spell out the details of the potential.

A well known general mechanism for the pair production of new particles is via the Drell-Yan process. Specifically, we can have
\beq
\label{eq;DYpsi}
q + \bar{q}\ra \gm^*\ra \psi^{+++} \psi^{---}.
\eeq
The decays of $\psi$ proceed as
\beq
\begin{split}
\label{eq:3+dec}
\psi^{+++}\ra W^+ +&S_1^{++}\\
%&\rotatebox[origin=c]{180}{$\Lsh$}\, \ell^+ + \ell^{\prime +}
&\rotatebox[origin=c]{90}{$\longleftarrow$}\\
&\ell^+ + \ell^{\prime +}.
\end{split}
\eeq
The final signature is a resonance of a same sign dilepton with a same sign $W$-boson.
In this reaction all the couplings are known with the only model dependence coming in the branching
ratio of $S_1\ra \ell \ell^\prime$.
An equally interesting reaction is
\beq
\label{eq:DYW}
u +\bar{d}\ra W^{+*}\ra \psi^{+++}+ S_1^{--},
\eeq
followed by the decay of $\psi^{+++}$ as in Eq.(\ref{eq:3+dec}) and a same opposite sign dilepton recoiling against it. Notice that none of the new charged states couples directly to quarks; hence,
the Drell-Yan mechanism is the best for their production.

High energy lepton colliders will be ideal probes for the new states, in particular if
we have a $ e^- e^-$ collider option. Such an advanced lepton collider is expected to operate with the center of mass (cm) energy in the multi-TeV range \cite{ale} and an exploratory luminosity of at least $10^{36} \mathrm{cm}^{-2} \mathrm{s}^{-1}$. Optimistically,
one can search for direct production of the doubly charged states as a dilepton resonance
via
\beq
\label{eephi}
e^- e^- \ra S_1^{--}\ra \ell^- \ell^{\prime -},
\eeq
where $ \ell,\ell^{\prime} = e,\mu,\tau$. Since $\bar{y}_\Phi \le 1$, this rate is not be suppressed.
If $M_1< \sqrt s$ with $s$  the cm energy, one will see a peak in the total cross section. For $M_1> \sqrt s$,
the cross section $e^- e^- \ra S^{*--} \ra \ell^- \ell^{\prime -}$ is
\beq
\label{eq: eeS}
\sigma =\frac{y_{ee}^2y_{\ell\ell^\prime}^2}{32 \pi} \frac{s}{(s-M_1^2)^2}.
\eeq
For the diagonal terms $\ell=\ell^\prime$, it is a factor of 2 larger. This cross section is
$O(40) \mathrm{fb}$ for $\sqrt{s}= 1\mtev$ if we set $y_{ee}=y_{\ell\ell^\prime}=e$.

 Similarly, one can consider the case of $e^- e^- \ra S^{*--}\ra W^- W^-$, which will be easier
 to search for in a lepton collider than a hadron collider. The reaction $e^- e^- \ra W^- W^-$ is the
 inverse of \NLDBD if the latter proceeds via a virtual W exchange. Thus, it provides a model independent  test of the two-W-boson mechanism for \NLDBD.
 Here, the W-boson pair is on shell.  This can be seen in the righthand
 diagram of Fig.~\ref{fig:0nubb}. Now the two electrons are incoming from the top and the
 two W-bosons are outgoing and decay into two jets each or $\ell \nu$ pair.
 This reaction must occur if \NLDBD were observed and proceeds via  the 2-W exchange. We refer this as inverse \NLDBD and was first discussed in
 \cite{LBN}. In our model, this is a s-channel
 process, hence any one of the W-bosons will have an isotropic scattering angle distribution. This contrasts with
the previous discussions on this reaction which were mainly focussed on
of probing heavy Majorana neutrino exchanges~\cite{LBN}. The latter has a characteristic
$t$  channel angular distribution that peaks at $\pi/2$. It is instructive to note that for a 1 TeV Majorana neutrino $N$ that mixes with $\nu_e$
with the mixing parameter $10^{-3}$, the cross section at high $s$ is $\sim 4.2$ fb. In passing we also note that similar probe reaction at hadron colliders such as the LHC using two W fusion to two same sign leptons have challenging backgrounds, see e.g. \cite{NPP}. On the positive side
searches at hadron colliders probe two units lepton number violation that are not both electrons.
Up till now the only constraints come from rare meson decays such as $K\ra \pi \mu \mu (\mu e)$
and $\mu^- \ra e^+$ conversion in nuclei.

\section{Conclusions}

Lepton number violation (LNV) is a crucial question in particle physics. It is intimately connected
to the question of the neutrino mass generation, which remains unknown despite tremendous progress in the experimental front in establishing neutrino oscillations. It is also widely believe that
it is violated by a small amount in the SM in terms of the small active neutrino Majorana mass. However,
this is far from being established. Observation of \NLDBD will then be the explicit demonstration of LNV independent of any model. There is now a world wide effort in improving the current experiment \cite{APPEC}.
The usual theoretical discussion begins with the assumption of long range exchanges of light
Majorana neutrinos as the dominant mechanism for \NLDBD. Here, we make the assumption that
\NLDBD proceeds mainly through short range interactions involving the two-W-boson exchange. We also
confine ourself to no new gauge interactions for the SM fermions. This turns out to be very restrictive
and new scalars with high $SU(2)$ representations can induce such decays. The new physics
also generates a small Majorana neutrino mass for $\nu_e$ that is insignificant for \NLDBD. While this is consistent with our hypothesis for \NLDBD but inconsistent with the
oscillation data. We propose to solve it by assuming that the light neutrinos have predominantly
Dirac masses and the small Majorana masses induced by the new scalars render them quasi-Dirac
particles. The splitting although small but may be detectable in the next generation of
neutrino oscillation experiments and/or neutrino telescope. This is a new connection between
\NLDBD and neutrino physics which is  yet to be studied in detail. Conversely the search for
evidence of pseudo-Dirac nature of neutrinos can shed light on the mechanism for \NLDBDs.

We also noted the new physics signals such as the high charged states that can be explored in hadron colliders.
In particular, we find that a high energy $e^-e^-$ will be very useful in testing the origin of LNV and complements
the \NLDBD studies.

\section*{ACKNOWLEDGMENTS}
This work was supported in part by National Center for Theoretical Sciences and
MoST  (MoST-107-2119-M-007-013-MY3).


\begin{thebibliography}{99}
\bibitem{DPR}
M.J.~Dolinski, A.W.P.~Poon, and W. Rodejohann,\ Rev. \ Nucl.\ Part. \ Sci. {\bf 69}, 219 (2019).
\bibitem{BG}
S.M.~Bilenky and C.~Giunti,\ Int. \ J. \ Mod. \ Phys. {\bf A30}, 1530001 (2015).
\bibitem{GDIK}
L.~Graf, F.E.~Deppisch, F.~Iachello, and J.~Kotila, \ Phys. \ Rev. {\bf 98}, 095023 (2019).
\bibitem{CDVGM}
V.~Cirigliano, W.~Dekens, J.~de Vries, M.L.~Graesser, and E.~Mereghetti, J. High Energy \ Phys. 12, 097 (2018).
\bibitem{PHKK99}
H.~P\"{a}s, M.~Hirsch, H.V.~Klapdor-Kleingrothaus, and S.~Kovalenko, \ Phys. \ Lett. {\bf B453}, 194 (1999).
\bibitem{PHKK01}
H.~P\"{a}s, M.~Hirsch, H.V.~Klapdor-Kleingrothaus, and S.~Kovalenko, \ Phys. \ Lett. {\bf B498}, 35 (1999).
\bibitem{SV}
J.~Schechter and J.W.F.~Valle, \ Phys. \ Rev. {\bf D25}, 774 (1982).
\bibitem{MLM}
M.~Duerr, M.~ Lindner and A.~Merle, J. High Energy \ Phys. 06, 091 (2011).
\bibitem{CGN}
C.S.~Chen, C.Q.~Geng, and J.N.~Ng, \ Phys. \ Rev. {\bf D75}, 053004 (2007).
\bibitem{CGNW}
C.S.~Chen, C.Q.~Geng, J.N.~Ng and J.M.S.~Wu,
  %``Testing radiative neutrino mass generation at the LHC,''
  JHEP {\bf 0708}, 022 (2007).

\bibitem{Atlasch}
M.~Aaboud {\it et.al.}, Atlas collaboration,\ Eur. \ Phys. \ J. {\bf C78}, 199 (2018).
\bibitem{CMSch}
CMS collaboration CMS-PAS-HIG-16-036 (2017).
\bibitem{CGH}
C.S.~Chen, C.Q.~Geng, D.~Huang and T.H.~Tsai, \ Phys. \ Rev. {\bf 87}, 077702 (2013).
\bibitem{KamL}
A.~Gando {\it {et al}} (KamLAND-Zen Collaboration), \ Phys. \ Rev. \ Lett. {\bf 110}, 062502 (2013).
\bibitem{cuore}
D.Q.~Adams {\it et. al}, CUORE Collaboration [arXiv:1912.10966] (2019).
\bibitem{gerda}
A.~Smolnikov, GERDA Collaboration , AIP \ Conf. \ Proc. {\bf 2165}, 020024 (2019).
\bibitem{nu2018}
Particle Data Group, \ Phys. \ Rev. {\bf  D98}, 030001 (2018).
\bibitem{CNW2}
W.F.~Chang, J.N.~Ng and J.M.S.~Wu, \ Phys. \ Rev. {\bf D80}, 113013 (2009).
\bibitem{wolf}
L.~Wolfenstein, \ Nucl. \ Phys. {\bf B186}, 147 (1981).
\bibitem{petcov}
S.~Petcov, \ Phys. \ Lett. {\bf B110}, 245 (1982).
\bibitem{dGHJ}
A.~de Gouv\^e a, W.C. Huang, and J.~Jenkins, \ Phys. \ Rev. {\bf D80}, 073007 (2009).
\bibitem{KL}
M.~Kobayashi and C.S.~Lim, \ Phys. \ Rev. {\bf D64}, 013003 (2001).
\bibitem{Pont}
B.~Pontecorvo, \ Sov. \ Phys. JETP {\bf 6}, 429 (1957); {\it ibid} {\bf 7}, 172 (1958).
\bibitem{MNS}
Z.~Maki, M.~Nakagawa and S.~Sakata, \ Prog. \ Theor. \ Phys. {\bf 28}, 870 (1962).
\bibitem{JUNO}
F.~An {\it et al.}(JUNO Collaboration),\ J. \ Phys. G {\bf 43}, 030401 (2016).
\bibitem{SP}
S.~Pakvasa, \ Nucl. \ Phys. \ Proc. \ Suppl. {\bf 137}, 295 (2004).
\bibitem{esma}
A.~Esmaili, \ Phys. \ Rev. {\bf D81} 013006 (2010).
\bibitem{Katrin}
M.Aker {\it et.al.}, KATRIN collaboration [arXiv:1909.06048] (2019).
\bibitem{proj8}
B.~Monreal and J.A.~Fromaggio, \ Phys. \ Rev. {\bf D80}, 051301 (2009).
\bibitem{CNW}
W.F.~Chang, J.N.~Ng and J.M.S.~Wu, \ Phys. \ Lett. {\bf B 730}, 373 (2014).
\bibitem{CN1}
W.F.~Chang and J.N.~Ng, \ Phys. \ Rev. {\bf D 90}, 065034 (2014).
\bibitem{CN2}
W.F.~Chang and J.N.~Ng, J.\ Cosmol. \ Astropart. \ Phys.  {\bf 07}, 027 (2016).
\bibitem{ale}
E.~Adli {\it et.al.},ALEGRO Collaboration, [arXiv:1901.10370] (2019).
\bibitem{LBN}
D.~London, G.~ B\a'elanger and J.N.~Ng, \ Phys. \ Lett. {\bf 188}, 155 (1987).
\bibitem{NPP}
J.N.~Ng, A. de la Puente and B.W.P.~Pan, J.\ High Energy \ Phys. 12, 172 (2015).
\bibitem{APPEC}
A.~Giuliani, APPEC Double Beta Decay Report, [arXiv:1910.04688] (2019).
\end{thebibliography}
\end{document}